\documentclass[aps,pra,superscriptaddress,twocolumn,longbibliography,amsmath,amssymb]{revtex4-2}

\usepackage{graphicx,cancel}
\usepackage{setspace,color,appendix,physics,wasysym,xcolor,xparse}
\usepackage{appendix}
\usepackage[english]{babel}
\usepackage{amsmath,amssymb,amsfonts,mathtools}
\usepackage[colorlinks = true, urlcolor = black, citecolor = blue, linkcolor = red]{hyperref}
\usepackage{lineno}
\usepackage{subfig}

\newcommand{\identity}{1\kern-0.25em\text{l}}

\begin{document}

\title{Continuous Variable Quantum Key Distribution channel emulator for the SPOQC mission}

\author{Emma Tien Hwai Medlock}
\email{ethm500@york.ac.uk}
\affiliation{School of Physics, Engineering \& Technology and York Centre for Quantum Technologies, University of York, YO10 5FT York, U.K.}
\author{Vinod N. Rao}
\affiliation{School of Physics, Engineering \& Technology and York Centre for Quantum Technologies, University of York, YO10 5FT York, U.K.}
\affiliation{Quantum Communications Hub, University of York, U.K.}
\author{Ry Render}
\affiliation{School of Physics, Engineering \& Technology and York Centre for Quantum Technologies, University of York, YO10 5FT York, U.K.}
\affiliation{Quantum Communications Hub, University of York, U.K.}
\author{Timothy Spiller}
\affiliation{School of Physics, Engineering \& Technology and York Centre for Quantum Technologies, University of York, YO10 5FT York, U.K.}
\affiliation{Quantum Communications Hub, University of York, U.K.}
\author{Rupesh Kumar}
\email{rupesh.kumar@york.ac.uk}
\affiliation{School of Physics, Engineering \& Technology and York Centre for Quantum Technologies, University of York, YO10 5FT York, U.K.}
\affiliation{Quantum Communications Hub, University of York, U.K.}

\begin{abstract} 
In a free space optical (FSO) communication link from satellite to ground, the losses in the channel will be dynamic. Thus, the characterization of the FSO channel is of great importance and this can be emulated in the lab to evaluate the realistic performance of a satellite payload. In this work, we introduce a novel optical channel emulator capable of replicating these dynamics, especially for Low Earth Orbit based CubeSats. We demonstrate its ability to accurately emulate a satellite-to-ground optical communications channel under various atmospheric turbulence strengths, satellite trajectories, and optical ground station parameters at a given optical wavelength of interest. Our satellite channel emulator was designed to test and benchmark the performance of the continuous variable quantum key distribution payload for the Satellite Platform for Optical Quantum Communications mission - an in-orbit demonstrator for the UK's Quantum Communication Hub, to be launched in early 2026.
\end{abstract}

\maketitle

\section{Introduction}
With the increasing demand for higher data rates, classical free space communications is moving from radio wave based communications to optical communications \cite{henniger2010introduction,sadiku2016free,agiwal2016next,chen2020vision,chowdhury20206g}. This is typically preformed with satellite-to-ground or inter-satellite communications as seen with companies such as SpaceX \cite{trichili2020roadmap, chaudhry2020free, chaudhry2022temporary}. Free space optical (FSO) communication links are desirable for their high data rates, ease of infrastructure implementation for international links \cite{chan2006free} and de-risking reliance on fiber optical cables for critical infrastructures. Quantum communication is also beginning to implement satellite-based FSO links, to provide information theoretical secure communication keys \cite{rarity2002ground,diamanti2015distributing}. In quantum communications, FSO links help to overcome the limitations of optical fiber links to achieve long and intercontinental distances \cite{dequal2021feasibility, vasylyev2019satellite, pirandola2021satellite}. Intercontinental distances could also be achieved with the use of quantum repeaters \cite{briegel1998quantum}, although this technology is not yet available \cite{azuma2023quantum}.

Since the launch of the first quantum key distribution (QKD) satellite Micius, launched by China in 2017 \cite{lu2022micius}, various nations have been working on national and international collaborative projects on satellite based QKD \cite{scott2020qeyssat, hiemstra2025european, knips2022qube, vergoossen2020spooqy, balakier2025high, jennewein2014nanoqey, oi2017cubesat, armengol2008quantum, sivasankaran2022cubesat}. This includes the UK's Satellite Platform for Optical Quantum Communications (SPOQC) mission, to be launched in early 2026 \cite{SPOQC}. The primary objective of this mission is a proof of principle demonstration of downlink based QKD for discrete and continuous variable payloads. Thus, we focus on demonstrating the emulation of such a channel in the lab.

The feasibility of different space-based FSO communication protocols can be tested by launching a satellite (which can be expensive and risky for preliminary protocol testing due to space weather \cite{chan2006free, pirjola2005space}), or by emulating the FSO channel in the laboratory. A satellite channel emulator enables testing of various communication protocols in the same system with minimal changes, flexibly and rapidly, and at lower development cost and risk. By using an FSO channel emulator, verification of the performance of a quantum payload (or classical) and its protocol can be performed during the Critical Design Review process of the mission. It can be used to test and compare various QKD protocols, and is also useful for benchmarking the quantum payload's secure key generation rate under realistic conditions \cite{roger2023real}. 

This paper introduces our FSO channel emulator for satellite-to-ground optical communication applications. It will emulate the dynamic channel losses as well as the beam aberrations expected to arise from the atmospheric turbulence effects. We show that our satellite channel emulator can be used to emulate an arbitrary satellite pass for any turbulence conditions along with various satellite/Optical Ground Station (OGS) parameters. This satellite channel emulator was specifically designed to test and benchmark the continuous variable (CV)-QKD payload for the SPOQC mission. In principle, it can emulate the channel for Discrete Variable (DV)-QKD based protocols or classical communications protocols as well. 

In this work we show the effects of various parameters, such as wavelength, transceiver apertures, satellite altitude, turbulence strength, etc. on the total loss of the channel. We also demonstrate the atmospheric aberration due to turbulence on the signal. We aim to emulate the channel for the CV-QKD protocol in the SPOQC mission. Additionally, we show the adaptability of the emulator to incorporate other channel conditions. The manuscript is presented as follows, Sec. \ref{theory} shows the channel loss calculations and channel conditions under our considerations. Sec. \ref{method} introduces the emulator, its experimental set-up and how each component emulates the various channel conditions. The results are shown and discussed in Sec. \ref{result}. Finally, this paper is concluded in Sec. \ref{conclusion}.

\section{Satellite-to-ground channel} \label{theory}

A satellite-to-ground FSO channel is highly dynamic. The light propagation through it experiences various effects that degrade signal integrity and contribute to signal loss and error. These include turbulence, beam diffraction, atmospheric attenuation, beam wandering, etc \cite{vasylyev2019satellite}. In this section, we will examine the effect of these sources on the FSO link in detail. 

\subsection{Atmospheric turbulence}

Atmospheric turbulence refers to the irregular, chaotic motion of air within the atmosphere. When the Sun heats the air, warm columns rise and then cool, causing them to sink back toward the Earth. This constant cycle generates turbulent eddies, each with varying refractive indices \cite{tyson2022principles}. As a beam of light passes through these eddies, it becomes perturbed. Turbulence strength is conventionally measured with the refractive index structure parameter $C_{N}^{2}$ and is defined by the mean square average of the refractive index $N$ between 2 points \cite{tyson2022principles, otoniel2015atmospheric}. The refractive index structure parameter was initially introduced by Hufnagel in 1964 and was later expended upon by Valley, Ulrich, Walters, Keunkel, Andrew and Phillips \cite{hufnagel1964modulation, valley1980isoplanatic, ulrich1988hufnagel, walters1981atmospheric, andrews2009near}. The refractive index structure parameter can be modeled in different ways. Here, the Hufnagel-Andrew-Phillips model was chosen, as it is more accurate at lower altitudes and becomes the conventionally used Hufnagel-Valley model at altitudes higher than $1~\text{km}$. For the Hufnagel-Andrew-Phillips model, the altitude dependent $C_{N}^{2}$ value is given by \cite{andrews2009near},

\begin{align}
C_N^2(h)=&M\bigg[0.00594\left(\frac{v_w}{27}\right)^2(10^{-5} h)^{10}\exp\left(\frac{-h}{1000}\right) \nonumber \\
&+2.7\times10^{-16}\exp\left({\frac{-h}{1500}}\right) +C_N^2(h_0)\left(\frac{h_0}{h}\right)^p \bigg] ,
\label{eqn:cn2}
\end{align}

where $h$ is the altitude, $v_w$ is the root-mean-square high altitude wind speed, $h_0$ is the height of the instrument above the ground, $C_N^2(h_0)$ is the average refractive index structure parameter at $h_0$, $M$ is the random background turbulence and $p$ is the power-law parameter depending on the time of day. For vertical propagating light, $C_N^2$ will be weighted based on altitude. Its value for a ground to satellite uplink channel, $I_0^{\text{up}}(\theta)$, and satellite to ground downlink channel, $I_0^{\text{down}}(\theta)$, are given by \cite{fante1980electromagnetic},

\begin{align}
&I_0^{\text{up}}(\theta) = \int_0^{z(\theta)}dx\bigg(1-\frac{x}{z(\theta)}\bigg) ^{5/3}C_N^2[h(x,\theta)], \nonumber \\
&I_0^{\text{down}}(\theta)=\int_0^{z(\theta)}dx\bigg(1-\frac{x}{z(\theta)}\bigg)^{5/3}C_N^2[h(z(\theta)-x,\theta)],
\label{eqn:I_0}
\end{align}

where $ z(\theta)$ is the propagation path length from the satellite to ground (or vice versa) and $\theta$ is the zenith angle. Using this equation, the turbulence effect on the beam can be calculated. 

Beam aberrations induced by the refractive index structure parameter can be represented with Zernike polynomials \cite{von1934beugungstheorie, born2013principles}. Zernike polynomials are used to model the spatial phase of light, representing wavefront aberrations \cite{noll1976Zernike}. The Zernike series contains all aberrations such as, piston, tilt, distortion, focus, astigmatism and spherical aberrations. One formalism for the Zernike series is in polar coordinates, with amplitude and phase $(\rho,\psi)$ respectively, and is defined by the following \cite{tyson2022principles},

\begin{align} \label{eq:zernike_series}
&\Phi(\rho ,\psi)=A_{00}+\frac{1}{\sqrt{2}}\sum_{n=2}^{\infty}A_{n0}R_{n}^{0}\bigg(\frac{\rho}{R^{'}}\bigg) \nonumber \\
&+\sum_{n=1}^{\infty}\sum_{m=1}^{n}[A_{nm}\cos m\psi+B_{nm}\sin m\psi]R_{n}^{m}\bigg(\frac{\rho}{R^{'}}\bigg),
\end{align}

where, $A_{nm}$ and $B_{nm}$ the Zernike coefficients and can be calculated from the power-series coefficients. The indices $n$ and $m$ denote the radial and azimuth order respectively, where $n$ and $m$ are always integral and satisfy $m \leq n,\text{ } n - |m| \in 2\mathbb{Z}$. $R^{'}$ is the radius over which the polynomials are defined, with the radial polynomials $R_{n}^{m}$ given by,

\begin{align} \label{eq:zernike_radial}
R_{n}^{\pm m}\bigg(\frac{\rho}{R^{'}}\bigg)=\sum_{s=0}^{\Delta_-}\frac{(-1)^{s}(n-s)!}{s!(\Delta_+ - s)!(\Delta_- -s)!}\bigg(\frac{\rho}{R^{'}}\bigg)^{n-2s}.
\end{align}

\begin{figure}
\centering
\includegraphics[width=\linewidth]{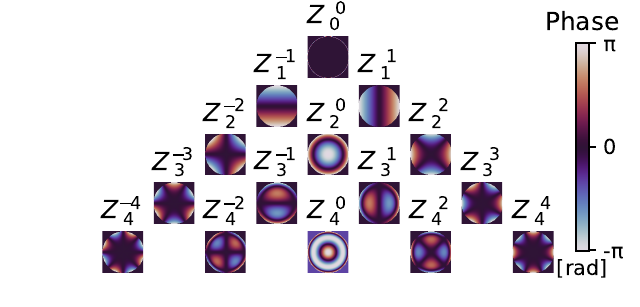}
\caption{Visual representation of the first 15 Zernike polynomials, generated using AO tools \cite{townson2019aotools}.}
\label{fig:zernike_series}
\end{figure}

Here, $s$ indexes the expansion and $\Delta_{\pm} = \frac{n\pm m}{2}$. Both Eq. (\ref{eq:zernike_series}) and (\ref{eq:zernike_radial}) contain mixtures of Seidel terms as shown in Tab. \ref{tab:seidel}. A visual representation of the first 15 Zernike polynomials can be seen in Fig. \ref{fig:zernike_series}.

\begin{table}[]
\centering
\begin{tabular}{|c|c|c|c|c|}
n & m & representation & description\\
\hline
0 & 0 & 1 & piston \\
1 & 1 & $\rho \cos \psi$ & tilt, distortion \\
2 & 0 & $\rho^{2}$ & focus, field curvature, sphere\\
2 & 2 & $\rho ^{2} \cos ^{2} \psi$ & astigmatism, cylinder\\
3 & 1 & $\rho ^{3} \cos \psi$ & coma \\
4 & 0 & $\rho ^{4}$ & spherical aberration
\end{tabular}
\caption{Seidel term representation}
\label{tab:seidel}
\end{table}

\subsection{Atmospheric attenuation and beam divergence} 

Signal attenuation is not only caused by absorption and scattering, but also by beam divergence. The loss from atmospheric molecular absorption can be neglected by carefully selecting the wavelength of the signal and we only consider the attenuation by scattering \cite{vasylyev2019satellite}. It is given as 

\begin{equation}
T_{atm}=e^{-\alpha_0g(\theta)},
\label{eqn:atten}
\end{equation}

where $\alpha_0$ is the extinction factor, and it depends on the scattering equally due to both aerosol distribution and Rayleigh scattering. With

\begin{equation}
g(\theta)=\int_0^{z(\theta)}e^{-h(x,\theta)/\bar{h}}dx,
\end{equation}

where $\Bar{h}=6600m$ and is a scaling factor and $h(x,\theta)$ is the altitude based off of the distance the beam has propagated by $x$ and the zenith angle $\theta$. 

Beam divergence is another source that contributes to signal loss when propagating over a free space channel. For a diffraction-limited beam propagated over a distance $z$, the transmittance is given by the following equation

\begin{equation}
T_{\text{dis}} = 2 \int_0^{a_r /2}\int_{d_r-x}^{d_r+x} \frac{2}{\pi \omega_{st} (\theta)^2}\exp(\frac{-2r^2}{\omega_{st}(\theta)^2})drdx,
\label{eqn:T_dis}
\end{equation}

with $d_r = \sqrt{d^2+(a_r/2-x)^2}$, $r$ is the radial distance from the center of the beam, $a_r$ is the receiver telescope aperture, $d$ is the displacement of the receiver telescope from the center of the beam which for this case can be assumed to be zero since only loss due to diffraction needs to be calculated and $\omega(\theta)_{st}$ is the turbulence dependent propagating beam width. With \cite{alda2003laser},
\begin{equation}
\omega_{st}(\theta)^2=\omega(\theta)^2+2\bigg(\frac{\lambda z(\theta)}{\pi r_0}\bigg)^2(1-\varphi)^2,
\label{eqn:omegast}
\end{equation}

where $\omega_{0}$ is the beam waist, $z$ is the propagation distance, $r_0$ is the Fried parameter, $\lambda$ is the wavelength of the signal, $\varphi=0.33\big(\frac{r_0}{\omega_0}\big)^{1/3}$ and $\omega(\theta)$ is the propagating beam width for a diffraction limited beam defined as follows,

\begin{equation} \label{eqn:omega(z)}
 \omega(\theta)=\omega_{0}\sqrt{1+\bigg(\frac{z(\theta)\lambda}{\pi\omega_{0}^2}\bigg)^2}. 
\end{equation}

This shows that loss due to beam divergence is also highly dependent on the wavelength of the propagating light.

\subsection{Pointing error and beam wandering}
Misalignment of the satellite pointing at the OGS causes signal losses, as seen Eq. (\ref{eqn:T_dis}) where the parameter $d$ represents this misalignment. This misalignment arises from two factors: (a) pointing error- the angular deviation between the signal beam from the satellite and the OGS receiver telescope aperture and (b) beam wandering due to turbulence. Both of these displace the beam center from the receiver telescope. The amount of displacement due to pointing error can be calculated using the following equation, 

\begin{equation}
d_{\text{Tr}}=a_z\tan{\bigg(\frac{t_{\text{err}}}{2}\bigg)},
\label{eqn:d_err}
\end{equation}
where $a_z$ is the altitude of the satellite and $t_{err}$ is the pointing error. Beam wandering due to turbulence is found with the displacement variance given by the following equation \cite{fante1975electromagnetic, yura1973short}, 

\begin{equation}
\sigma_{TB}^2=\frac{0.1337\lambda^2z^2}{\omega_0^{1/3}r_0^{5/3}},
\label{eqn:sig_TB}
\end{equation}

where $\lambda$ is the wavelength, $z$ is the propagation path length, $\omega_0$ is the beam waist and $r_0$ is the Fried parameter that depends on the turbulence strength.\\

\section{Experimental set-up and methodology} \label{method}
Given the FSO channel parameters discussed in Sec. \ref{theory}, the channel can be simulated and emulated. In order to emulate the satellite-to-ground FSO channel, three main components were used. A variable optical attenuator (VOA) to emulate loss from beam divergence and atmospheric attenuation, a fine steering mirror (FSM) to emulate the pointing error and turbulence induced beam wandering, and a deformable mirror (DM) to emulate turbulence. Details on how these components were used to emulate the specific channel conditions are shown in the following Sec. \ref{method:defrac}, \ref{method:point} and \ref{method:atmos}.

\begin{figure}
\centering
\subfloat[\centering Satellite Channel Emulator set-up diagram]{\includegraphics[width=\linewidth]{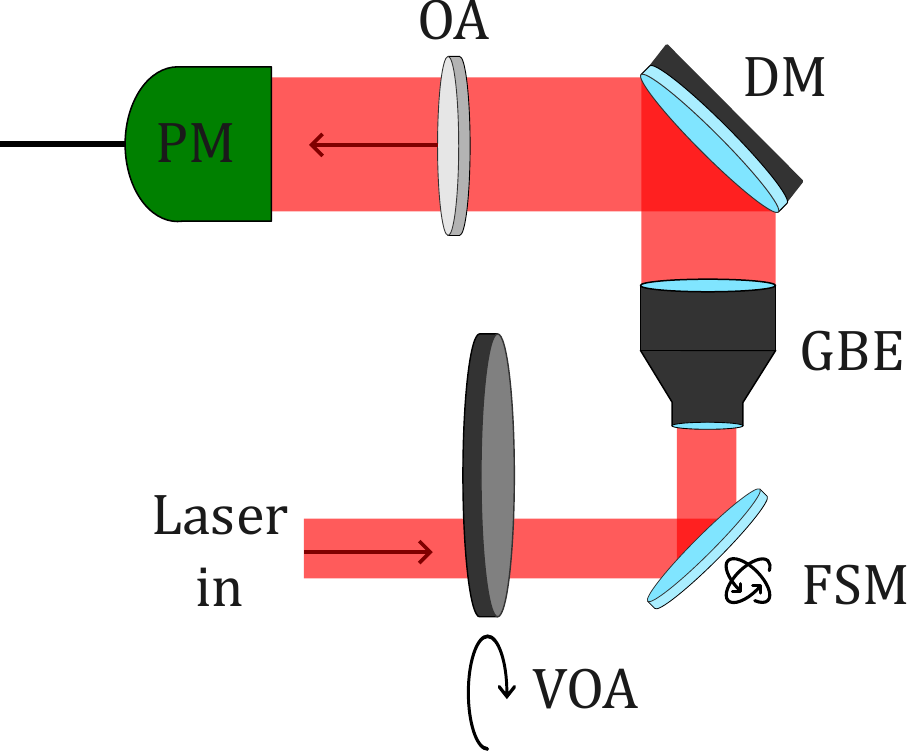} \label{fig:set-up1}}
\qquad
\subfloat[\centering Satellite Channel Emulator set-up photograph]
{\includegraphics[width=\linewidth]{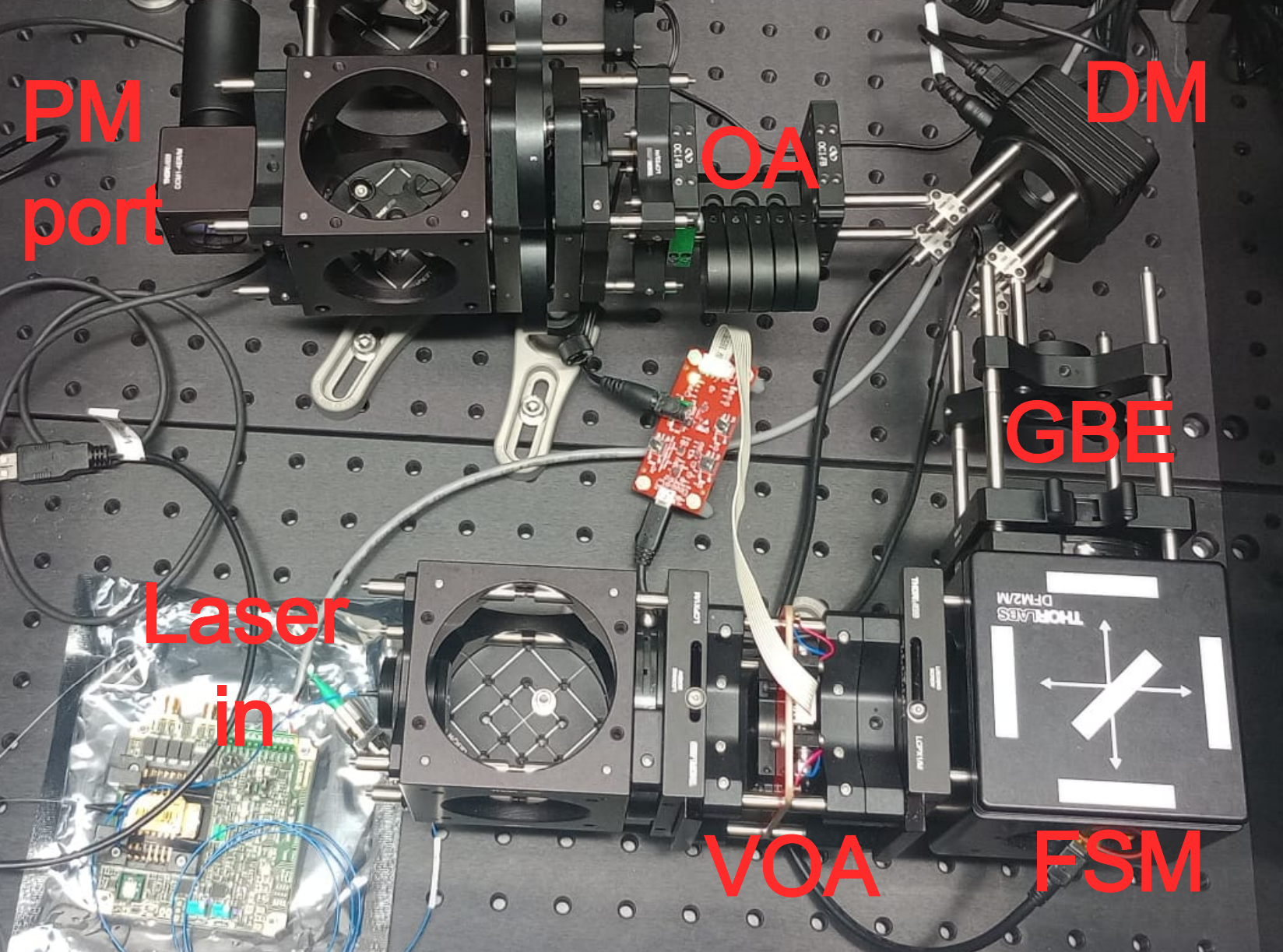} \label{fig:set-up2}}
\caption{Experimental set-up of the satellite-to-ground channel emulator as a diagram in (a) and photograph in (b). Here the VOA emulates the loss due to diffraction and atmospheric attenuation, the FSM emulates the pointing error and beam wander due to turbulence, GBE expands the beam, DM emulates the beam aberrations due to atmospheric turbulence affects, OA (optical attenuator) for addition beam attenuation and PM (power meter) measures the losses the signal experiences through this channel.}
\label{fig:set-up}
\end{figure}

The emulator is shown in Fig. \ref{fig:set-up}. A $2~\text{mm}$ wide collimated beam from a CW laser passes through the VOA, which adds losses due to divergence and atmospheric attenuation to the beam. The FSM, which wanders randomly in a given range, induces random displacement to the beam to emulate pointing error and beam wandering. The beam is then expanded to $1~\text{cm}$ with a Galilean beam expander (GBE) and passes to the DM of 1 cm aperture. This beam expansion is critical for DM to induce desired aberrations on the input beam. A phase screen, depending on the turbulence strength is prepared and sent to the DM to impart the relevant beam aberrations on the beam. Additional fixed losses can be induced with optical attenuators (OA) if necessary. A power meter (PM) measures the signal loss at the exit aperture of the emulator set-up. Error to the beam such as change in time of arrival of signal pluses and wavelength fluctuation form the Doppler effect and microgravity and change in linear polarization due to the attitude of the satellite with respect to a receiver are not considered in this set-up. These errors will only effect specific communications protocols. For inter-satellite links, the Doppler effect and micro gravity becomes relevant, for local local oscillator (LLO) based CV-QKD the Doppler effect becomes relevant and for polarization encoding based DV-QKD protocols the change in linear polarization becomes relevant. These specific errors are not within the scope of this work, there are plans to incorporate them in future iterations of the emulator.
 
As seen from Eq. (\ref{eqn:I_0}), (\ref{eqn:atten}), (\ref{eqn:T_dis}) and (\ref{eqn:d_err}), the parameters depending on the direction of beam propagation- uplink or downlink- are those dependent on atmospheric turbulence. Therefore, the developed set-up can interchangeably emulate both uplink and downlink channels with appropriate phase screens on the DM and beam wandering range setting on the FSM. The setting parameters in the emulator can accommodate satellite altitude, transmitter and receiver apertures, signal wavelength, atmospheric conditions, etc. In short, it has the capabilities to emulate an arbitrary satellite channel ranging from LEO to geostationary (GEO) and beyond. And in addition inter-satellite optical communication links can be emulated as well. A detailed functional descriptions of the components of the emulator are given below in Subsections \ref{method:defrac}, \ref{method:point} and \ref{method:atmos}.

\subsection{Channel attenuation} \label{method:defrac}

For the emulation of atmospheric attenuation and loss from beam divergence, a variable optical attenuator was used (25FS04DV.4, neutral density filter from Newport with variable optical density from $0-4~\text{OD}$). The amount of attenuation at any given point was controlled and changed by rotation of the VOA with the use of a motorised rotation mount (ELL14K Rotation Mount with Resonant Piezoelectric Motors from Thorlabs). From Eq. (\ref{eqn:atten}) and (\ref{eqn:T_dis}), the losses due to atmospheric attenuation and diffraction will be dependent on propagation path length (and therefore also zenith angle) and the wavelength of the laser used for this FSO communication. Additionally, as seen in Eq. (\ref{eqn:T_dis}), the loss due to beam divergence is also dependent on the apertures of the satellite and OGS telescopes. Based on the various parameters that contribute, the variable optical attenuator is set to mimic the overall optical link loss.

The rate at which the rotation mount can update the loss is around $1.8~\text{Hz}$, which corresponds to an angular displacement of $0.6^{\circ}$ at the $\pm30^{\circ}$ zenith angle and $0.35^{\circ}$ at zenith for a satellite at an altitude of $700~\text{km}$. We consider the variation in loss to be negligible within this period.

\begin{figure}
\centering
\subfloat[\centering VOA loss for single satellite pass]{\includegraphics[width=\linewidth]{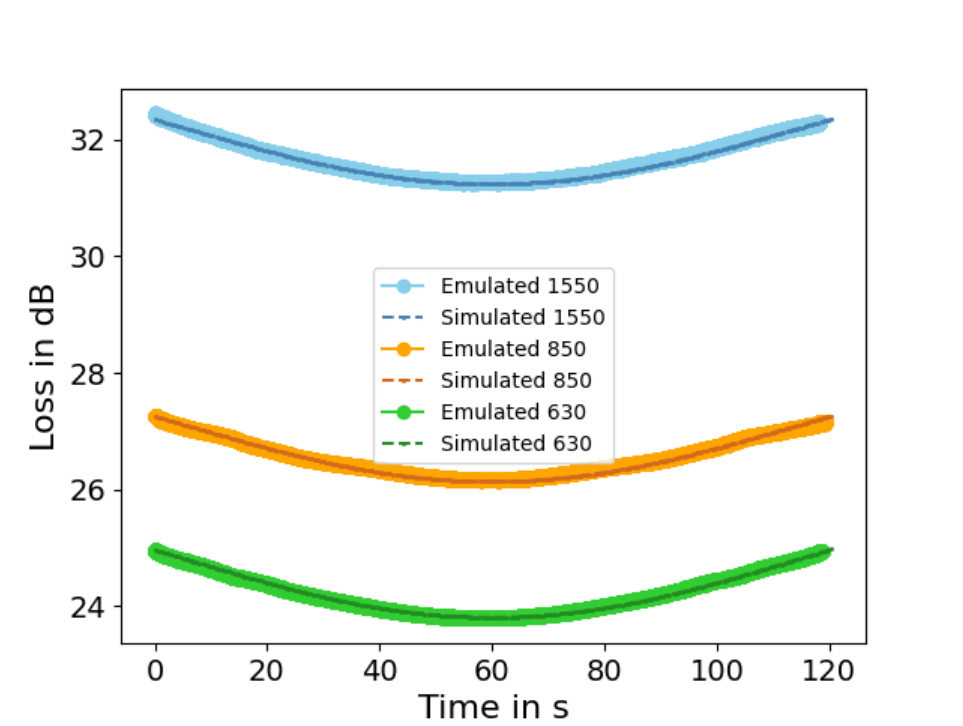} \label{fig:voa_loss1}}
\qquad
\subfloat[\centering Probability distribution of VOA emulated loss for single satellite pass]
{\includegraphics[width=\linewidth]{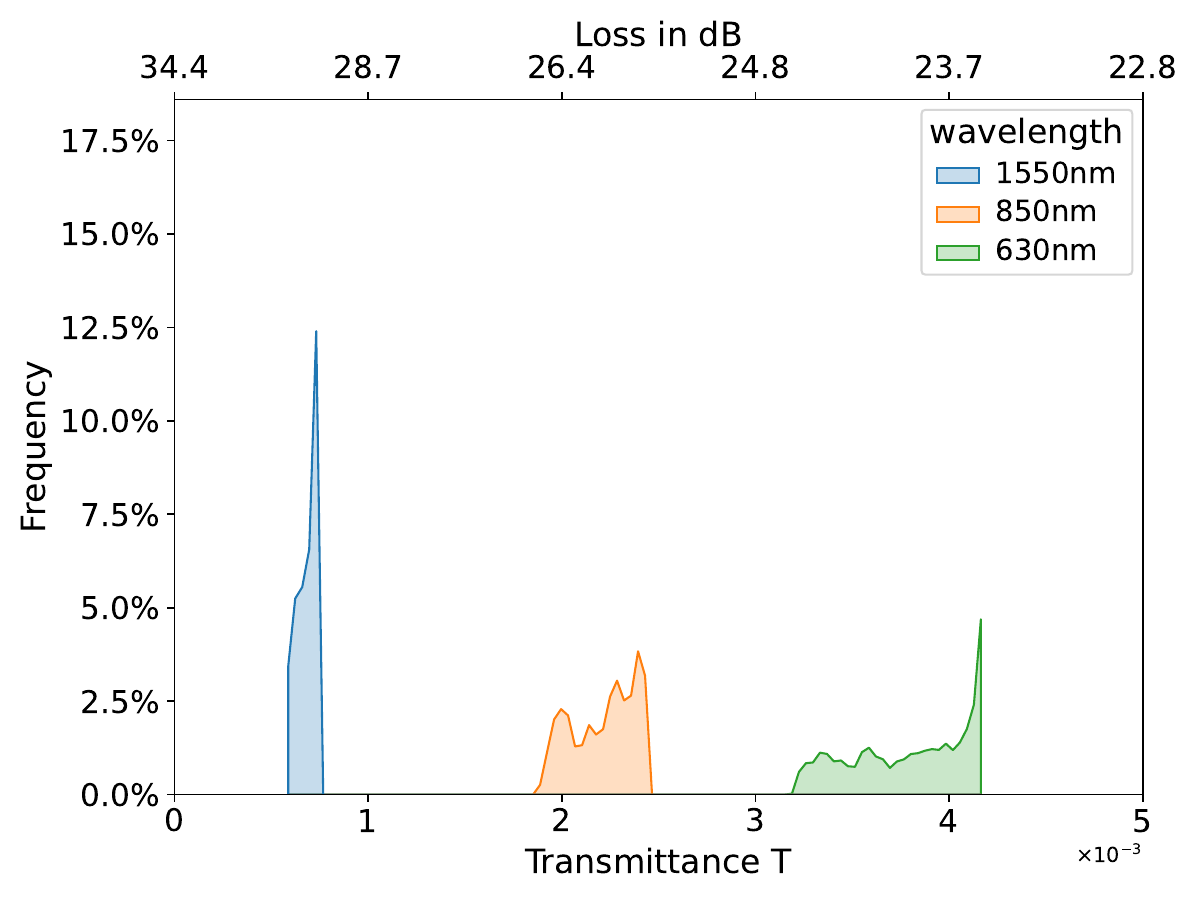} \label{fig:voa_loss2}}
\caption{Loss at the VOA given a satellite at altitude of $700~\text{km}$, telescope aperture at the satellite of $8~\text{cm}$ and OGS telescope aperture of $60~\text{cm}$ with $30\%$ obstruction. Emulated for three different wavelengths, $1550~\text{nm}$, $850~\text{nm}$ and $630~\text{nm}$. (a) comparison emulated and simulated loss from VOA and (b) probability distributions of emulated loss from the VOA.}
\label{fig:voa_loss}
\end{figure}

The simulated theoretical loss compared to the emulated loss by the VOA is shown in Fig. \ref{fig:voa_loss1}. Here, the loss was calculated and emulated for a satellite at an altitude of $700~\text{km}$, telescope aperture at the satellite of $8~\text{cm}$, OGS telescope aperture of $60~\text{cm}$ with $30\%$ obstruction and for three different wavelengths. As seen in the figure, the VOA can accurately emulate the loss from divergence and atmospheric attenuation for various wavelengths.

The probability distribution of the emulated loss from the VOA can be seen in Fig. \ref{fig:voa_loss2}, this matches the probability distribution for these losses in Dequal et. al. \cite{dequal2021feasibility}. The figure clearly shows a greater spread in loss at shorter wavelengths, but higher loss at longer wavelengths. This can be seen from Eq.(\ref{eqn:T_dis}) and (\ref{eqn:omega(z)}), as longer wavelengths will be affected more by divergence than shorter wavelengths.

\subsection{Signal displacement} \label{method:point}

To emulate the loss from pointing error and beam wandering, a fine steering mirror (TMAU2520X45Y20 mems electrostatic micro mirror from Sercalo) was used. This FSM has a maximum range of $\pm4.5^{\circ}$ along the x-axis and $\pm2^{\circ}$ along the y-axis. The FSM was programmed to mimic the beam wandering movements (due to pointing error and turbulence) by moving randomly. This FSM has a repetition rate of $1~\text{kHz}$. The typical time scale for beam wandering due to turbulence effects and pointing error is slow and on the scale of $~10~\text{ms}-100~\text{ms}$ \cite{pirandola2021satellite, bourgoin2013comprehensive}. The repetition rate of the FSM is enough to emulate both the pointing error and beam wandering. The random movements of the FSM were programmed within the range of the maximum displacement of the beam from the center of the telescope. The amount of displacement and therefore the range of the FSM can be found with Eq. (\ref{eqn:d_err}) and (\ref{eqn:sig_TB}). 

\begin{figure}
\centering
\subfloat[\centering FSM emulated loss for single satellite pass]
{\includegraphics[width=\linewidth]{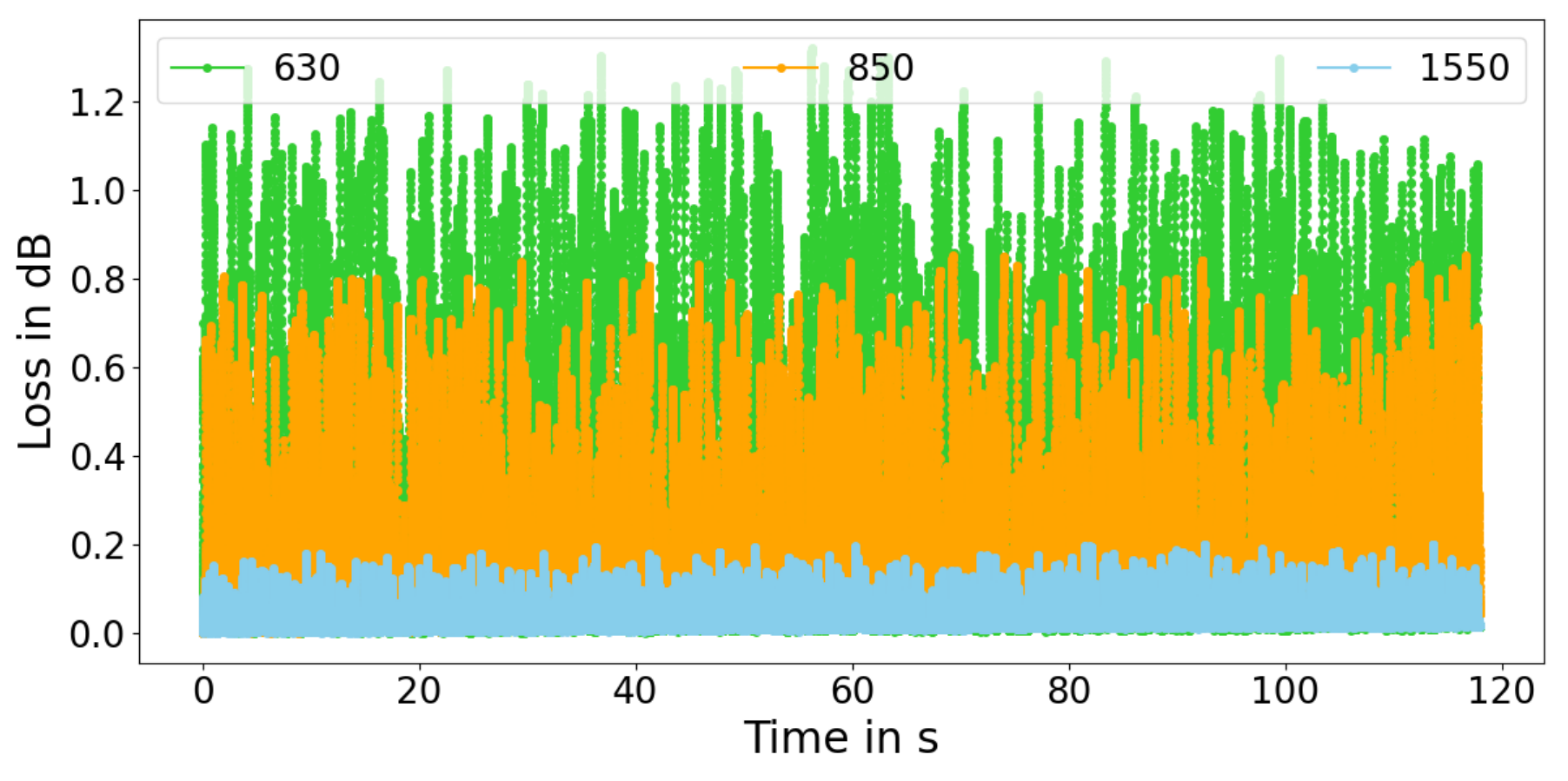} \label{fig:FSM1}}
\qquad
\subfloat[\centering Probability distribution of FSM emulated loss for single satellite pass]
{\includegraphics[width=\linewidth]{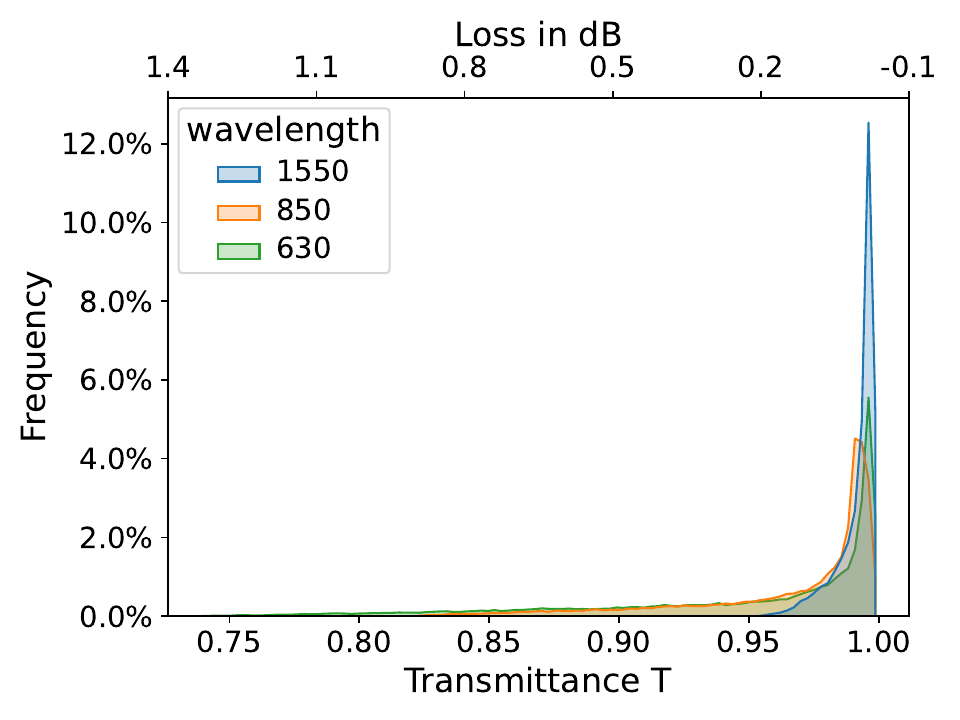} \label{fig:FSM2}}
\caption{(a) FSM emulated loss, with input parameters as follows: a satellite at an altitude of $700~\text{km}$, a ground level turbulence strength of $C_N^2(h_0)=10^{-12}~\text{m}^{-2/3}$, a pointing error of $4~\mu\text{rad}$ and a OGS receiver telescope of $60~\text{cm}$ with $30\%$ obstruction. This was emulated for three different wavelengths, $1550~\text{nm}$, $850~\text{nm}$ and $630~\text{nm}$. (b) is the portability distribution of FSM emulated loss.}
\label{fig:FSM}
\end{figure}

The Fig. \ref{fig:FSM1} shows the loss due to beam wandering and pointing error at different wavelengths, from a satellite at an altitude of $700~\text{km}$, under strong turbulence conditions with $C_N^2(h_0)=10^{-12}~\text{m}^{-2/3}$, a pointing error from the satellite of $4~\mu\text{rad}$, a satellite telescope aperture of $8~\text{cm}$ and an OGS receiver telescope of $60~\text{cm}$ with $30\%$ obstruction. 

Fig. \ref{fig:FSM2} shows the probability distribution of the FSM loss data which follows a Weibull distribution and matches the distribution for pointing error and beam wandering found in Dequal et. al. \cite{dequal2021feasibility}. This distribution arises from the geometry of the gaussian beam intensity profile and the geometry of the receiver \cite{dequal2021feasibility}. The maximum combined loss given these parameters for pointing error and beam wandering due to turbulence is around $0.2~\text{dB}$, $0.9~\text{dB}$ and $1.3~\text{dB}$ for $1550~\text{nm}$, $850~\text{nm}$ and $630~\text{nm}$ respectively, see Eq. (\ref{eqn:T_dis}), (\ref{eqn:d_err}) and (\ref{eqn:sig_TB}) for a given wavelength, satellite and OGS parameters. Again, there is a greater spread in loss for shorter wavelengths. But unlike for the VOA emulated losses, there is a greater overall loss for shorter wavelengths. Longer wavelengths will be less affected by atmospheric turbulence and the loss due to pointing error is lessened since the beam spread will be larger due to diffraction. Therefore, being displaced from the center of the beam due to pointing error has a less adverse effect loss, as evident from Eq. (\ref{eqn:T_dis}). 

For different input parameters, the FSM bounds can be changed, and the FSM can emulate the losses for other satellite, OGS and wavelength configurations. The FSM can emulate any desired configuration and is only limited by the maximum angle range of the FSM. Therefore, with the use of an FSM to emulate signal displacement, both uplink and downlink for GEO or LEO as well as inter-satellite links can be emulated. 

\subsection{Turbulence emulation} \label{method:atmos}

To emulate turbulence effects, a deformable mirror was used (DMH40-P01 High-Stroke Deformable Mirror from Thorlabs). This instrument uses the first $15$ Zernike polynomials as inputs (see Fig. \ref{fig:zernike_series}) to generate the desired beam aberrations. This DM cannot emulate tip/tilt, as it does not have actuators for this. The FSM was used for this function instead, as it can be used to emulate turbulence-based beam wandering. For the tip/tilt functionality, i.e. beam wandering due to turbulence, the refractive index structure parameter was used instead of the Zernike mode to determine the amount of beam wander. As seen in Sec. \ref{method:point}, Eq. (\ref{eqn:sig_TB}) was used. 

The beam aberrations and Zernike polynomials were calculated and generated using the python tool set AO tools \cite{townson2019aotools}. This tool set can generate a random phase screen for a given turbulence strength and receiver aperture size, from which it can then calculate the Zernike polynomials that create this aberration. From Eq. (\ref{eqn:I_0}) the refractive index structure parameter can be calculated as a function of propagation path length, which in turn depends on the zenith angle of the satellite with respect to the OGS \cite{cakaj2011range}. Therefore, the refractive index structure parameter can be calculated at a given zenith angle for uplink as well as downlink channels. The DM can then iterate through multiple phase screens at a maximum repetition rate of $1~\text{kHz}$. The time-scale of scintillation is highly dependent on the wind speed and is in the order of milliseconds \cite{dravins1997atmospheric, osborn2015atmospheric}, which is within the repetition rate of the DM.

\begin{figure}
\centering
\subfloat[\centering Single shot phase screen]
{\includegraphics[width=\linewidth]{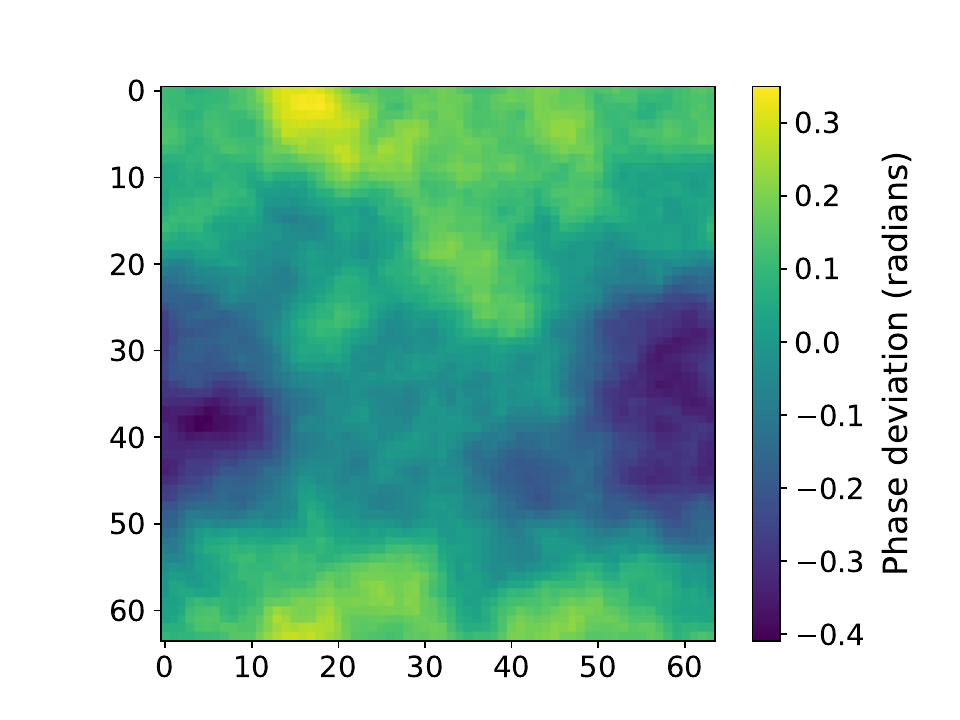} \label{fig:turb1}}
\qquad
\subfloat[\centering Phase screen Zernike polynomials vs Noll's turbulence theory]
{\includegraphics[width=\linewidth]{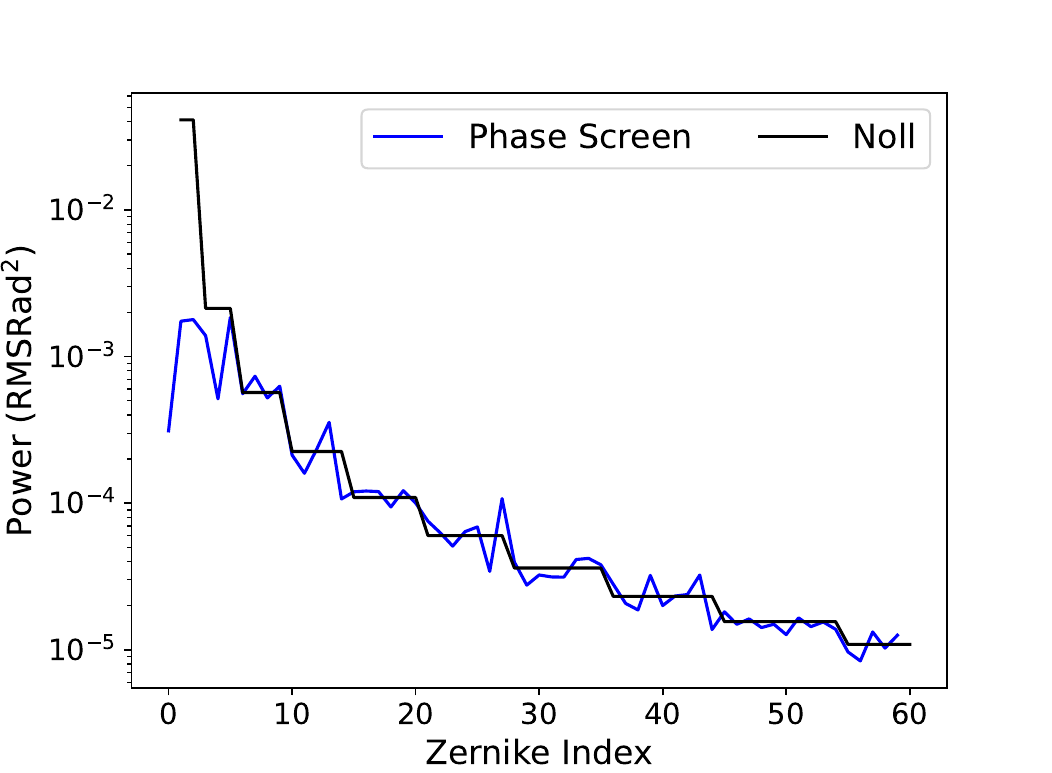} \label{fig:turb2}}
\caption{Single shot phase screen at the zenith given ground level turbulence of $C_N^2(h_0)=10^{-12}~\text{m}^{-2/3}$, $1550~\text{nm}$ light and $60~\text{cm}$ with $30\%$ obstruction OGS telescope aperture. (a) the randomly generated phase screen and (b) the Zernike polynomials of said phase screen compared to Noll's turbulence theory \cite{noll1976Zernike}. The DM can take the first 15 Zernike polynomials as an input to replicate these phase screens.}
\label{fig:turb}
\end{figure}

In order to generate the phase screens with AO tools, first the turbulence strength needs to be calculated using Eq. (\ref{eqn:I_0}). Since the refractive index structure parameter will be changing with the satellite overpass, the finite Kolmogorov phase screen function in AO tools was used to generate the phase screens. The following function was used -- ``\textit{phasescreen.ft\_phase\_screen(r0, nx\_size, pxl\_scale, L0, l0)}''. Where \textit{r0} is the Fried parameter, \textit{nx\_size} is the amount of pixels, \textit{pxl\_scale }is the pixel scale and depends on the chosen diameter of the beam or receiver aperture, \textit{L0} is the outer scale of turbulence and \textit{l0} is the inner scale of turbulence. The inner scale of turbulence is the size of the smallest turbulence eddies and the outer scale of turbulence is the size of the largest turbulence eddies \cite{tyson2022principles}. Here, standard values for inner and outer scales for turbulence eddies were chosen. The generation methods for finite Kolmogorov phase screens are given in \cite{kolmogorov1991local}. The Zerike modes of the generated phase screens can then be found using the following AO tools function, \textit{aotools.zernikeArray(n\_zerns, nx\_size)}, which decomposes the phase screen into its Zernike polynomials using methods from \cite{noll1976Zernike}. Here \textit{n\_zerns} is the number of Zernike polynomials calculated. Only the first $15$ Zernike modes were generated, as the DM can not use any of the higher modes. This should not impair the emulation of the atmospheric turbulence effect on the communications, as the higher modes have comparatively lower effect on the aberrations \cite{von1934beugungstheorie}. Fig. \ref{fig:turb1} shows a single phase screen at the zenith for a $1550~\text{nm}$ wavelength, a receiver aperture of $60~\text{cm}$ with $30\%$ obstruction and a strong ground level turbulence of $C_N^2(h_0)=10^{-12}~\text{m}^{-2/3}$. The Zernike polynomial values calculated from the phase screen can be compared to Noll's theory of turbulence from \cite{noll1976Zernike}. The result is shown in Fig. \ref{fig:turb2}, which is in good agreement with the finite phase screen given in \cite{townson2019aotools}. 

\begin{figure}
\centering
\includegraphics[width=\linewidth]{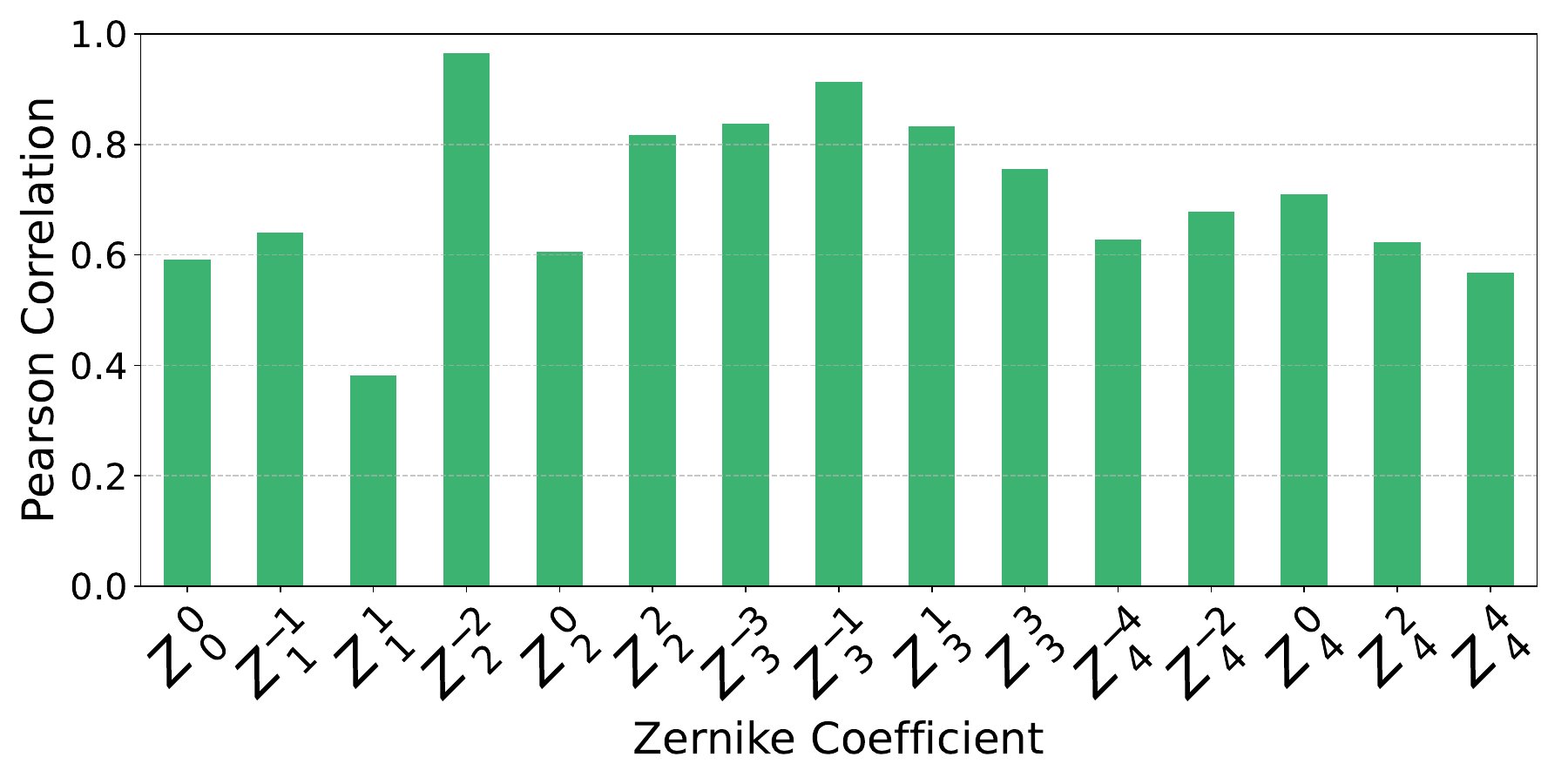}
\caption{Correlations between the Zernike modes inputted to the DM and the Zernike modes measured with the Shack-Hartmann wavefront sensor.}
\label{fig:wfs_data}
\end{figure}

To verify the DM is functioning as intended with the phase screens, a Shack-Hartmann wavefront sensor (WFS20-K2/M Shack-Hartmann Wavefront Sensor from Thorlabs with a $300~\mu\text{m}$ Pitch AR-Coated Microlens Array) was used to measure the beam aberration from the DM. For this test the DM was set to a $45^\circ$ angle, this is the same angle the DM is set to in the satellite channel emulator (see Fig. \ref{fig:set-up}). The correlations between the input phase screen to the DM and measured beam aberration as per the phase screens, for a given Zernike mode, can be seen in Fig. \ref{fig:wfs_data}. This figure shows a good correlation between the the input Zernike modes and the measured Zernike modes, although there are some slightly smaller correlations, as the wavefront sensor axis misalignment from the DM (measurements are taken off-axis), which is to be expected.

\begin{figure}
\centering
\subfloat[\centering DM emulated loss for single satellite pass]
{\includegraphics[width=\linewidth]{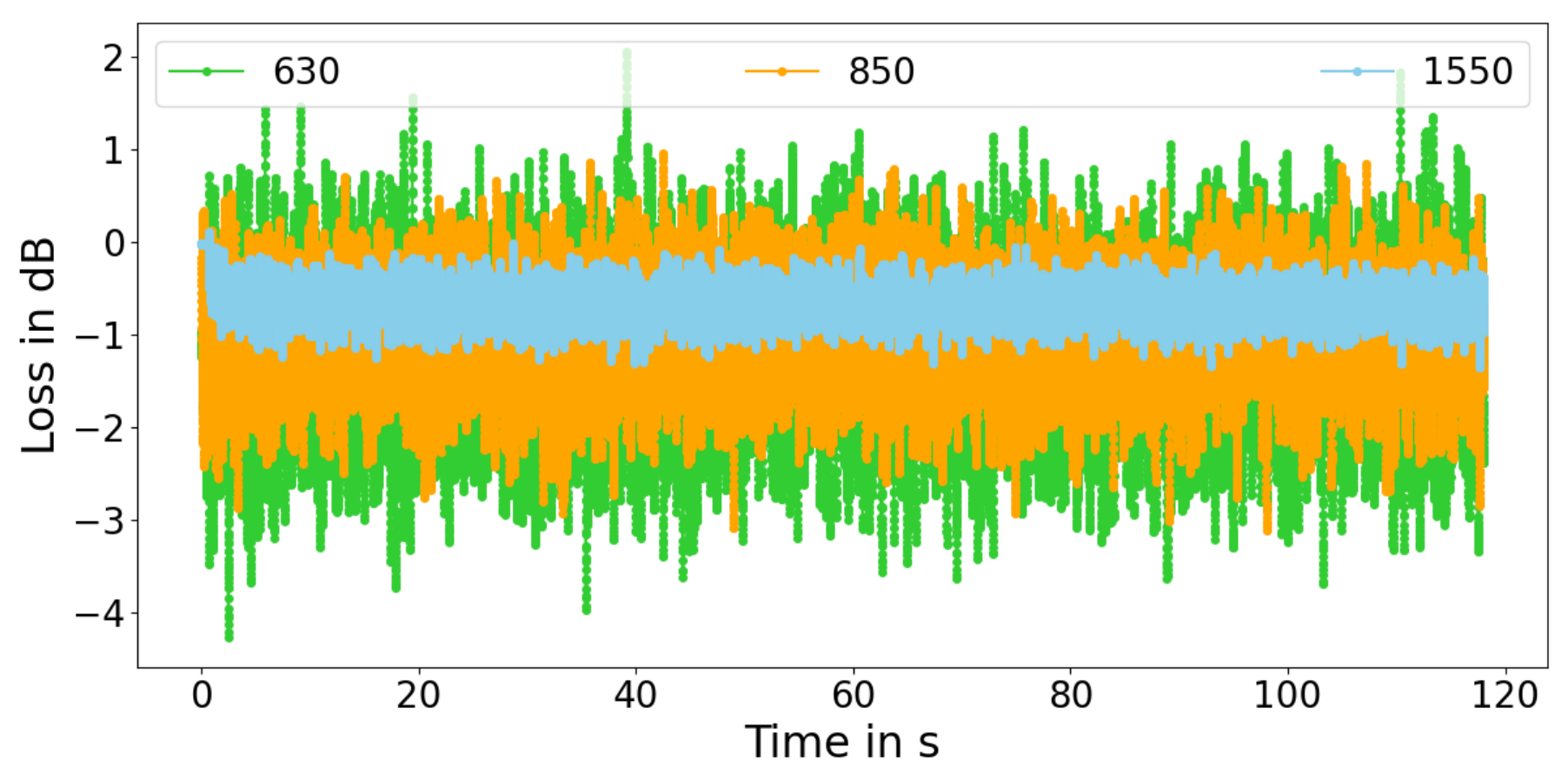} \label{fig:turb_loss1}}
\qquad
\subfloat[\centering Probability distribution of DM emulated loss for single satellite pass]
{\includegraphics[width=\linewidth]{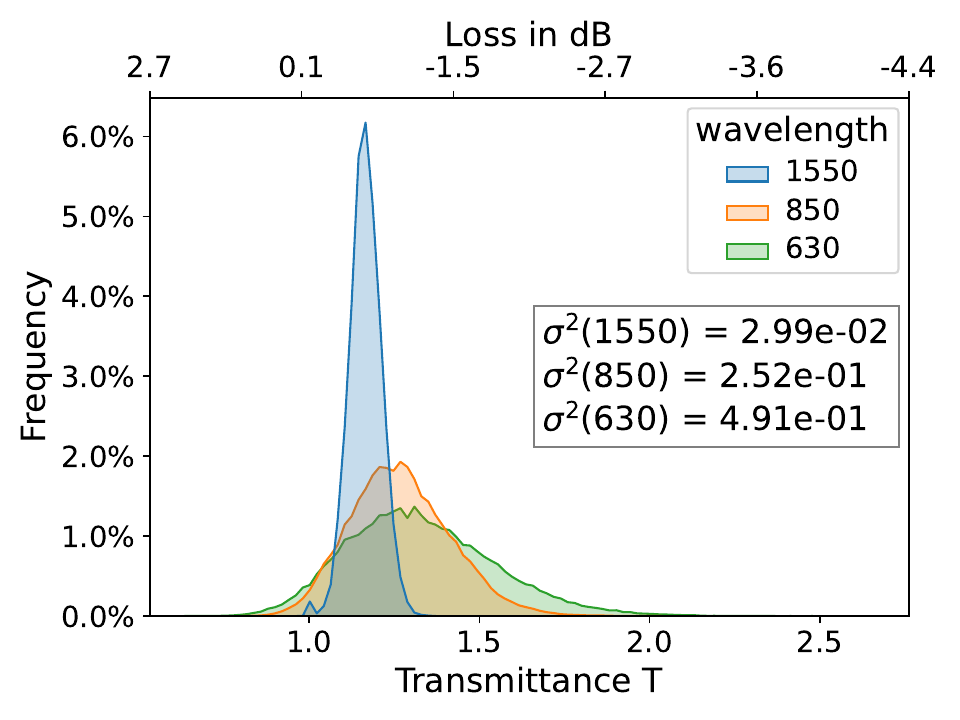} \label{fig:turb_loss2}}
\caption{(a) full satellite pass DM emulated loss for three wavelengths, $1550~\text{nm}$, $850~\text{nm}$ and $630~\text{nm}$. The input parameters were a transmitter aperture of $8~\text{cm}$, a receiver aperture of $60~\text{cm}$ with $30\%$ obstruction, with the satellite at an elevation of $700~\text{km}$ and a ground level turbulence of $C_N^2(h_0)=10^{-12}~\text{m}^{-2/3}$. (b) is the portability distribution of the DM emulated loss. The loss variance for a given wavelength can be seen in the legend.}
\label{fig:turb_loss}
\end{figure}

The loss from turbulence emulated for the satellite overpass can be seen in Fig. \ref{fig:turb_loss1} for the three different wavelengths given a transmitter aperture of $8~\text{cm}$, a receiver aperture of $60~\text{cm}$ with $30\%$ obstruction, with the satellite at an elevation of $700~\text{km}$ and a ground level turbulence of $C_N^2(h_0)=10^{-12}~\text{m}^{-2/3}$. As seen in Eq. (\ref{eqn:I_0}), the turbulence strength is dependent on propagation path length and therefore the zenith angle of the satellite. Given the parameters chosen for this emulation although, the change in truculence strength due to angle change has a negligible effect on the change in loss and as seen in Fig. \ref{fig:turb_loss1}, the loss variance does not change noticeably with the satellite pass. 

The probability distribution for this loss can be seen in Fig. \ref{fig:turb_loss2}, which matches the lognormal distribution for scintillation loss found in \cite{lyke2009probability,giggenbach2017scintillation}. As seen in the figure, the loss variance increases with decreasing wavelength. This trend matches the theory, where the loss variance can be found from \cite{tyson2022principles, fante1975electromagnetic, churnside1991aperture} for a given wavelength, satellite, OGS telescope and turbulence parameters. Shorter wavelengths will be more affected by turbulence compared with longer wavelengths. Here, the loss variance is $0.03~\text{dB}$, $0.25~\text{dB}$ and $0.5~\text{dB}$ for $1550~\text{nm}$, $850~\text{nm}$ and $630~\text{nm}$ respectively. 

\section{Results and discussion} \label{result}

\begin{figure}
\centering
\subfloat[\centering Total loss for single satellite pass]
{\includegraphics[width=\linewidth]{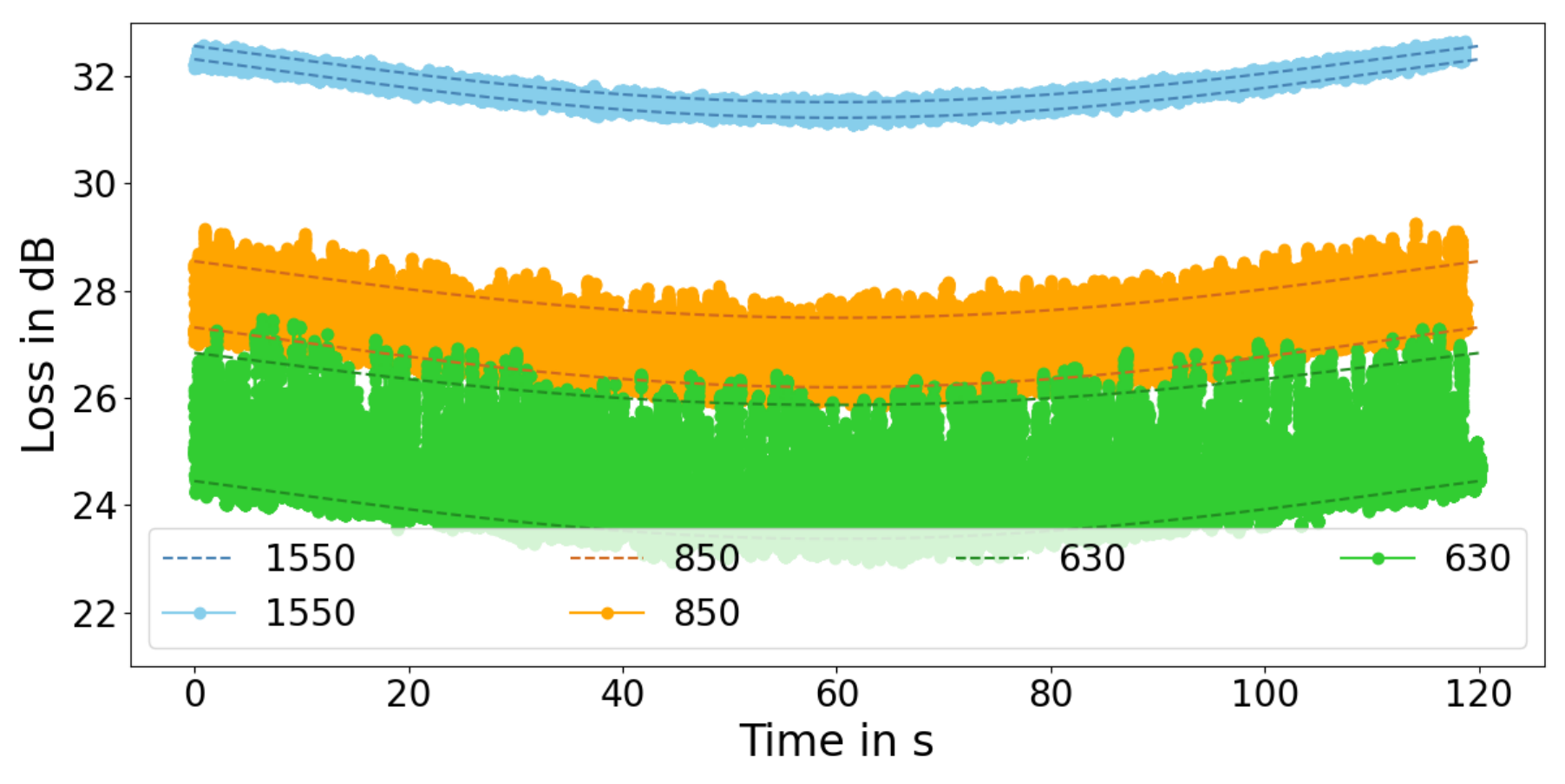} \label{fig:total_loss1}}
\qquad
\subfloat[\centering Probability distribution for total emulated loss for single satellite pass]
{\includegraphics[width=\linewidth]{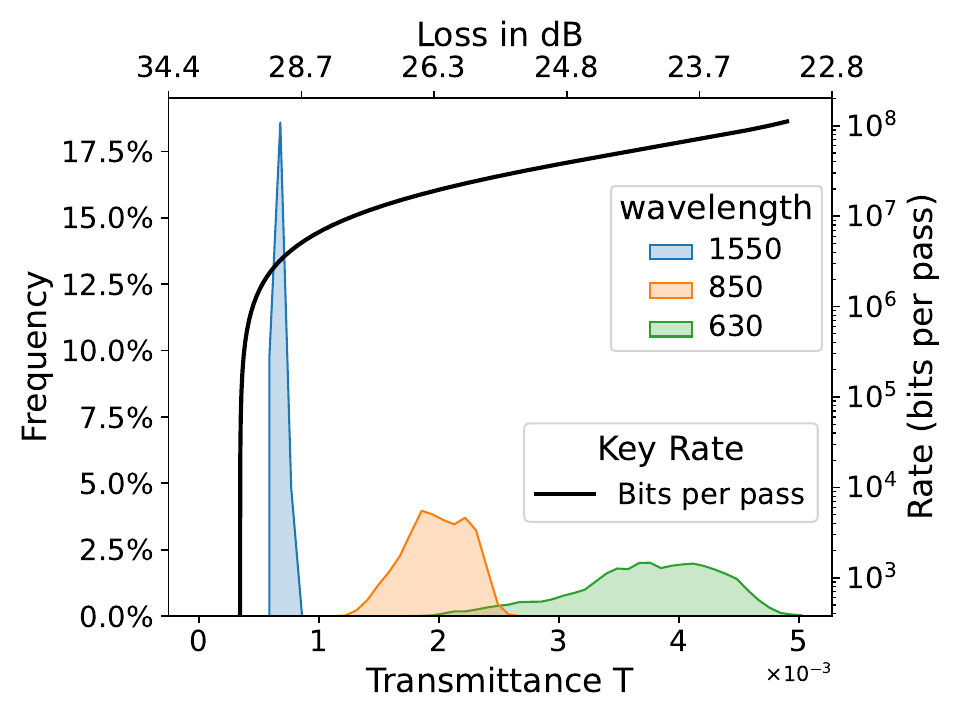} \label{fig:total_loss2}}
\caption{(a) emulated loss data compared to the simulated bounds for a single satellite pass. Dashed lines are the simulated bounds and the points are the measured emulated loss. This was done for $1550~\text{nm}$, $850~\text{nm}$ and $630~\text{nm}$. (b) the probability distributions for the loss data, as well as the achievable secret key per pass. Here the average secret key bits per pass are $3.1\times10^7$, $1.9\times10^8$ and $3.6\times10^8$ for $1550~\text{nm}$, $850~\text{nm}$ and $630~\text{nm}$ respectfully. With the following input parameters: transmitter aperture of $8~\text{cm}$, a receiver aperture of $60~\text{cm}$ with $30\%$ obstruction, a pointing error at the satellite of $4~\mu\text{rad}$, a large area free space detector, with the satellite at an elevation of $700~\text{km}$ and a ground level turbulence of $C_N^2(h_0)=10^{-12}~\text{m}^{-2/3}$.}
\label{fig:total_loss}
\end{figure}

For a full satellite pass channel emulation, multiple components in the set-up need to be synchronized, and the parameters of the satellite, OGS, turbulence and wavelength are inputted. Fig. \ref{fig:total_loss1} shows the total loss in the satellite to ground FSO channel under strong turbulence, loss from atmospheric attenuation \& beam divergence, and pointing \& beam wandering effects at wavelengths $1550~\text{nm}$, $850~\text{nm}$ and $630~\text{nm}$. The figure shows the total loss measured by the power meter after the light passes through the emulator compared to the simulated loss bounds. 

Fig. \ref{fig:total_loss2} shows the respective probability distributions of the full satellite pass loss data for the given wavelengths, as well as the achievable secret key per pass (key rate calculation can be seen in the Appendix Sec. \ref{Appendix_A}). These probability distributions can be seen as constructive patterns from the probability distributions seen in Fig. \ref{fig:voa_loss2} from the VOA, Fig. \ref{fig:FSM2} from the FSM and Fig. \ref{fig:turb_loss2} from the DM. Here, the shorter wavelengths have a comparatively lower loss but have higher loss variance. According to Eq. (\ref{eqn:omega(z)}), shorter wavelengths result in a smaller propagating beam width. This leads to less loss from beam divergence when compared to longer wavelengths. Since beam divergence is a major contributor to overall loss, shorter wavelengths will experience less total loss. Although, as seen in the previous sections, the shorter wavelengths will be more affected by atmospheric turbulence, hence the variance in the loss will be larger. This can also be seen in the spread of the probability distributions for the different wavelengths, where for $630~\text{nm}$ the shape is smoother compared to longer wavelengths, as the effects from turbulence (see FSM Sec. \ref{method:point} and DM losses Sec. \ref{method:atmos}) are smoother. The loss probability distribution from the FSM is a Weibull distribution (as seen in Fig. \ref{fig:FSM2} and the DM has a lognormal distribution (as seen in Fig. \ref{fig:turb_loss2}). Both of these probability distributions will have a larger effect on the loss distribution of the full satellite channel emulation when shorter wavelengths are used. At $1550~\text{nm}$, the loss variance is smaller as turbulence has a much smaller effect. This is reflected by the shape of the probability distribution; here, the distribution is more jagged, as the main contributing factor to this shape is from the VOA loss (see Fig. \ref{fig:voa_loss2}. 

One can note the absence of polarization drift and Doppler effect generators from the emulator set-up. Both of these effects are irrelevant to the CV-QKD payload design. The CV-QKD payload uses a transmitting local oscillator (TLO) based implementation, in which both the QKD signal and a strong reference signal (the local oscillator) for the QKD signal measurement at the receiver are co-propagating through the FSO channel. Therefore, both signals remain in coherence with respect to each other. The polarization of both signals is also selected to be invariant to satellite orientation. Those features will be added to the emulator in the future to incorporate other QKD schemes, such as LLO based CV-QKD, polarization encoding-based DV-QKD, etc. The transmitted LO pulse also serves as the clock for time synchronization.

Although this emulator focuses on the channel characterization for CV-QKD in the SPOQC mission, it can be expanded to emulate channels for uplink, inter-satellite and GEO based FSO links. Full secret key generation with the use of the emulator for the SPOQC mission CV-QKD payload is in the works. In principle, the satellite channel emulator can also be used for various other CV- and DV-based protocols (including entanglement based protocols). This will be demonstrated in future work.

\section{Conclusion} \label{conclusion}
This work demonstrates a novel satellite-to-ground channel emulator for free space optical communication applications. We demonstrated that the developed set-up can emulate a satellite channel for arbitrary signal wavelengths under higher turbulence conditions and various satellite/OGS telescope parameters. The developed experimental set-up finds application in emulating channel conditions for the upcoming SPOQC mission for demonstrating QKD from space. This satellite channel emulator can now be used to test, benchmark, and verify the CV-QKD payload, constructed by the University of York team, for the SPOQC mission. The emulator is also useful for testing other QKD systems, such as DV-QKD and other classical FSO systems as well. The emulator can accommodate beam polarization alignment, Doppler effect compensation, etc, during the near future up-gradation activities. This emulator was specifically designed and tested to emulate the channel conditions, and therefore the losses from the receiver (homodyne/heterodyne detector for CV-QKD) are not emulated. 

\textbf{Funding:} Engineering and Physical Sciences Research Council (EP/T001011/1); University of York.

\textbf{Acknowledgment:} The authors acknowledge the funding support from EPSRC Quantum Communications Hub (Grant number EP/T001011/1). E.T.H.M thanks the School of Physics, Engineering and Technology, University of York for the PhD funding.

\textbf{Disclosures:} The authors declare no conflicts of interest

\textbf{Data Availability Statement:} The data supporting this research is openly available from the research data repository of the University of York at https://doi.org/10.15124/6e75c578-8fe2-4f6f-84a1-8fcddbbbac34.

\section{Appendix A: CVQKD key rate} \label{Appendix_A}

We consider the Gaussian-modulated coherent-state (GMCS) CVQKD protocol with reverse reconciliation under
collective attacks, assuming a trusted homodyne detector at Bob and a passive
eavesdropper constrained by line-of-sight (LoS) geometry \cite{ghalaii2023satellite}. The asymptotic secret
key rate is given by $K = \beta I_{AB} - \chi_{BE}^{\mathrm{LoS}}$, where $\beta$
denotes the reconciliation efficiency, $I_{AB}$ is the classical mutual
information between Alice and Bob, and $\chi_{BE}^{\mathrm{LoS}}$ is the Holevo
information between Bob and a LoS-restricted Eve.

For homodyne detection at Bob, the mutual information reads
$I_{AB} = \tfrac{1}{2}\log_2[(V + \chi_{\mathrm{tot}})/(1 +
\chi_{\mathrm{tot}})]$, where $V = V_A + 1$ is the variance of Alice's Gaussian
modulation. The total noise referred to the
channel input is $\chi_{\mathrm{tot}} = \chi_{\mathrm{line}} +
\chi_{\mathrm{hom}}/T$, with the channel-added noise given by
$\chi_{\mathrm{line}} = (1 - T)/T + \xi$ and the trusted homodyne detection noise
$\chi_{\mathrm{hom}} = (1 - \eta)/\eta + v_{\mathrm{el}}$, where $T$ is the
Alice--Bob channel transmittance, $\xi$ is the excess noise referred to the
channel input, $\eta$ is the detection efficiency, and $v_{\mathrm{el}}$ denotes
the electronic noise.

In the passive LoS-restricted eavesdropping model, Eve cannot access the entire
channel loss but only a geometrically limited fraction $T_E \le 1 - T$ of the
optical mode.  The Holevo information $\chi_{BE}^{\mathrm{LoS}}$ can be written as $\chi_{BE}^{\mathrm{LoS}} =
\sum_{i=1}^{2} G[(\lambda_i^{E} - 1)/2] - \sum_{j=3}^{5} G[(\lambda_j^{E} - 1)/2]$,
where the bosonic entropic function is $G(x) = (x+1)\log_2(x+1) - x\log_2 x$. Here, the eigen values
 $\lambda_{1,2}^{E} = \sqrt{\{A_E \pm \sqrt{A_E^2 - 4B_E}\}/2}$, with
$A_E = V^2(1 - 2T_E) + 2T_E + T_E^2(V + \chi_{\mathrm{line}})^2$ and
$B_E = T_E^2(V\chi_{\mathrm{line}} + 1)^2$.  The eigen values $\lambda_{3,4}^{E} = \sqrt{\{C_E \pm \sqrt{C_E^2 - 4D_E}\}/2}$, where
$C_E = [A_E\chi_{\mathrm{hom}} + V\sqrt{B_E} + T_E(V +
\chi_{\mathrm{line}})]/[T_E(V + \chi_{\mathrm{tot}})]$ and
$D_E = [\sqrt{B_E}(V + \sqrt{B_E}\chi_{\mathrm{hom}})]/[T_E(V +
\chi_{\mathrm{tot}})]$. And $\lambda_5^{E} = 1$. For plotting the key rate in Fig. \ref{fig:total_loss2} we have used the following parameters: $V_A$ = X; $\beta$ = X ; $v_{ele}$ = X; $\eta$ = X and $T_E$ = X, for the CVQKD system at clock rate 2 MHz during 120 seconds of the satellite transit time. The chosen parameters for key rate graph in Fig. \ref{fig:total_loss2} are as follows; excess noise $\xi_{\text{tot}}=0.003$, reconciliation efficiency $\beta=0.98$, detector efficiency of $\eta_{d}=0.8$, electronic noise $v_{\text{el}}=0$, optimal modulation variance $V_{\text{opt}}=300~\text{SNU}$, telescope coupling transmissivity $\eta_{T}=0.4$ and Alice-Eve channel transmittivity $\eta_{AE}=0.01$.

\bibliography{Bibliography}

\end{document}